\documentclass[11pt]{article}

\usepackage{amssymb,epsfig,fullpage}

\usepackage{subfigure}
\usepackage{graphicx}

\begin{document}

\title{Analysis of symmetries in models of multi-strain infections}
\author{K.B. Blyuss\thanks{Corresponding author. Email: k.blyuss@sussex.ac.uk}\\\\
Department of Mathematics, University of Sussex,\\Brighton, BN1 9QH, United Kingdom}

\maketitle

\maketitle

\begin{abstract}
In mathematical studies of the dynamics of multi-strain diseases caused by antigenically diverse pathogens, there is a substantial interest in analytical insights. Using the example of a generic model of multi-strain diseases with cross-immunity between strains, we show that a significant understanding of the stability of steady states and possible dynamical behaviours can be achieved when the symmetry of interactions between strains is taken into account. Techniques of equivariant bifurcation theory allow one to identify the type of possible symmetry-breaking Hopf bifurcation, as well as to classify different periodic solutions in terms of their spatial and temporal symmetries. The approach is also illustrated on other models of multi-strain diseases, where the same methodology provides a systematic understanding of bifurcation scenarios and periodic behaviours. The results of the analysis are quite generic, and have wider implications for understanding the dynamics of a large class of models of multi-strain diseases.
\end{abstract}

\section{Introduction}

In the analysis of infections with multiple strains simultaneously co-circulating in a population, an important role is played by antigenic diversity, where hosts can be infected multiple times with antigenically different strains of the same parasite, which allows the parasite to maintain its presence in the host population (Craig and Scherf 2003, Lipsitch and O'Hagan 2007). Major examples of pathogens employing antigenic diversity as a strategy of immune escape include malaria (Gupta et al 1994, Recker et al 2004), meningitis (Gupta and Anderson 1998, Gupta et al. 1996), dengue (Gog and Grenfell 2002, Recker et al 2009), and influenza (Earn et al 2002, Ferguson et al 2003, Smith et al 1999). From the perspective of interactions between different strains, one can distinguish between two major types of strain interactions: {\it ecological interference} where a host infected with one strain is removed from the population susceptible to other strains (Levin et al 2004, Rohani et al 2003), and {\it immunological interference}, where infection with one strain may confer partial or full immunity to other strains (Gupta and Anderson 1998) or lead to enhancement of susceptibility or transmissibility of other strains, as is the case for dengue (Recker et al 2009) and HPV (Elbasha and Galvani 2005). The underlying mechanism of cross-immunity is generic for all pathogens: an infection with one strain of a pathogen elicits a lasting immune memory protecting the host against infections with other immunologically related strains.

In terms of analysis of the dynamics of multi-strain diseases, in the last twenty years a significant number of mathematical models have been put forward that aim to explore and explain different aspect of interactions between multiple strains. In terms of implementation, one can divide these models into agent- or individual-based models and equation-based models. For the first class of models, pathogen strains are treated as individuals interacting according to some prescribed rules (Buckee et al. 2004, Buckee and Gupta 2010, Cisternas et al 2004, Ferguson et al 2003, Sasaki and Haraguchi 2000, Tria et al 2005), which allows for efficient stochastic representation of immunological interactions but does provide an intuition arising from analytical tractability. The second class of models provides two alternative treatments of cross-immunity between strains, known as history-based and status-based approaches. In history-based models, the hosts are grouped according to what strains of a pathogen they have already been infected with, and transitions between different compartments, which corresponds to infection with other strains, occur at rates depending on the strength of cross-protection between strains (Andreasen et al 1996, Andreasen et al 1997, Castillo-Chavez et al 1989, Gomes et al 2002, Gupta et al 1998, Gupta et al 1996, Lin et al 1999). On the other hand, in status-based models the hosts are classified not based on their previous exposures to individual strains but rather by their immune status, i.e. the set of strains to which a given host is immune (Gog and Grenfell 2002, Gog and Swinton 2002, Kryazhimsky et al 2007). Once in a particular immune compartment, upon infection with a new strain individuals move to other immune compartments at rates determined by the probabilities of acquiring cross-immunity against other strains. In this approach, partial cross-immunity can make some hosts become completely immune whilst other hosts will not gain immunity from the same exposure - this is known as {\it polarized immunity} (Gog and Grenfell 2002) and is equivalent to an alternative formulation used in the analysis of effects of vaccination (Smith et a 1984).

Since different strains of a pathogen form as a result of some common genetic process, they inherit immunological characteristics associated with this process. A convenient tool quantifying the degree of immunological relatedness between different strains arising from their antigenic structure is the {\it antigenic distance} between strains, which can take into account antigenic structure as determined by the configuration of surface proteins, as well as the difference in antibodies elicited in response to infection with another genotype (Gupta et al 2006, Smith et al 1999, Smith et al 2004). Conventionally, one assumes that the larger is the antigenic distance between two strains, the smaller is the level of cross-immunity between them. In mathematical models of multi-strain diseases, one of the effective ways to include antigenic distance is to use a multi-locus system (Gupta et al 1998, Gupta et al 1996), where each strain is represented by a sequence of $n$ loci with $m$ alleles in each locus, thus resulting in a discrete antigenic space (some authors have considered similar set-up in a continuous one-dimensional antigenic space (Adams and Sasaki 2007, Andreasen et al 1997, Gog and Grenfell 2002, Gomes et al 2002). In this approach, for any two given strains, the number of locations at which their sequences are identical determines their immunological relatedness, which is taken as a proxy measure of cross-immunity (Calvez et al 2005, Cobey and Pascual 2011, Ferguson and Andreasen 2002, Gupta et al 1998, Minaev and Ferguson 2009, Tria et al 2005). Alternatively, it is possible to map each genotype to a point in antigenic space (Koelle et al 2006, Recker et al 2007) and then separately introduce a function that determines the strength of cross-immunity between strains based on their antigenic distance (Adams and Sasaki 2007, Andreasen 1997, Gog and Grenfell 2002, Gomes et al 2002). 

Whilst significant progress has been made in the analysis of generic features of multi-strain models and possible types of dynamics they are able to exhibit, the effects of symmetry, which is present in many of the models, have remained largely unexplored. Andreasen et al (1997) have considered a multi-strain epidemic model with partial cross-immunity between strains. They analysed stability of the boundary equilibria representing symmetric steady states with only immunologically unrelated strains present, and also showed that the internal endemic equilibrium can undergo Hopf bifurcation giving rise to stable periodic oscillations. Furthermore, these authors also demonstrated how this periodic orbit can disappear in a global bifurcation involving a homoclinic orbit through a two-strain equilibrium. This work was later extended to a system of three linear-chain strains (Lin et al 1999), and again the existence of sustained oscillations arising from a Hopf bifurcation of internal endemic equilibrium was shown. Dawes and Gog (2002) have considered a generalised model of an SIR dynamics with four co-circulating strains and studied possible bifurcations leading to the appearance of periodic behaviour by performing bifurcation unfolding in the regime when the basic reproductive number very slightly exceeds unity. More recently, Chan and Yu (2013a,b) have used groupoid formalism to analyse symmetric dynamics in models of antigenic variation and multi-strain dynamics , and they have also demonstrated the emergence of steady state clustering as a result of symmetry properties of the system. Blyuss (2013) has investigated symmetry properties in a model of antigenic variation in malaria (see also Blyuss and Gupta (2009) for analysis of other related dynamical features), and Blyuss and Kyrychko (2012) have extended this analysis to study the effects of immune delay on symmetric dynamics.

In this paper we use the techniques of equivariant bifurcation theory to systematically study stability of steady states and classification of different types of periodic behaviour in a multi-strain model. Using a classical multi-locus model of Gupta et al (1998) as an example, we will illustrate how the symmetry in the interactions between strains can provide a handle on understanding steady states and their stability, as well as the emergence of symmetry-breaking periodic solutions.
The outline of this paper is as follows. In the next section we introduce the specific model to be used for analysis of symmetries in models of multi-strain diseases and discuss its basic properties. Section 3 contains the analysis of steady states and their stability with account for underlying symmetry of the model. In Sect. 4 different types of dynamical behaviours in the model are investigated and classified in terms of their symmetries. Section 5 illustrates how a similar methodology can be used for studying other types of multi-strain models. The paper concludes in Sect. 6 with discussion of results and future outlook.

\section{Mathematical model}

In order to study the effects of symmetry on dynamics in multi-strain models, we consider a multi-locus model proposed by Gupta et al (1998). In this model, $z_i(t)$ denotes a proportion of population who are immune to strain $i$, i.e. those who have been or are currently infected with the strain $i$, $y_i(t)$ is the fraction of population who are currently infectious with the strain $i$, and $w_i(t)$ is the proportion of individuals who have been infected (or are currently infected) by a strain antigenically related to the strain $i$, including $i$ itself (with $1\leq i\leq N$). The model equations can then be written as
\begin{equation}\label{GFAsys}
\begin{array}{l}
\displaystyle{\frac{dy_i}{dt}=\lambda_i[1-\gamma w_i-(1-\gamma)z_i]-(\mu+\sigma)y_i,}\\\\
\displaystyle{\frac{dz_i}{dt}=\lambda_i(1-z_i)-\mu z_i,}\\\\
\displaystyle{\frac{dw_i}{dt}=(1-w_i)\sum_{j\sim i}\lambda_j-\mu w_i,}
\end{array}
\end{equation}
where $\lambda_i$ is the force of infection with strain $i$ defined as $\lambda_i=\beta y_i$, where $\beta$ is the transmission rate assumed to be the same for all strains, $1/\mu$ and $1/\sigma$ are the average host life expectancy and the average period of infectiousness, respectively, and $\gamma$ $(0\leq\gamma\leq 1)$ is the cross-immunity, giving the reduction in transmission probability conferred by previous infection with one strain.
\begin{figure}
\hspace{4cm}
\epsfig{file=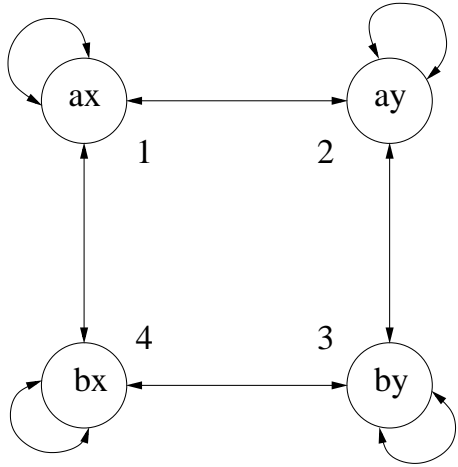,width=6cm}
\caption{Map of antigenic interactions between different strains in the two locus-two allele system.}\label{strains}
\end{figure}
In terms of disease transmission, the population is assumed to be randomly mixed, and upon recovery from infection with a particular strain, the immunity to that strain is lifelong. To characterize strains and their immunological interactions, each strain is described by a sequence of antigens consisting of $N_{L}$ loci, with $n_{k}$, $1\leq k\leq N_L$, alleles at each locus, so that $\Pi_{k=1}^{N_L}n_k=N$. In system (\ref{GFAsys}), expression $j\sim i$ refers to all strains $j$ sharing alleles with strain $i$. In the simplest non-trivial case of a two locus-two allele system represented by alleles $a$ and $b$ at one locus, and $x$ and $y$ at the other, we have a system of four antigenically distinct strains as shown in Fig.~\ref{strains}. A simple but justifiable assumption about such system is that as a consequence of immune selection, infection, for instance, with strain $ay$ will have a negative impact on transmission of strains $ax$ and $by$ but will have no impact on transmission of the strain $bx$, as they are completely immunologically distinct (Gupta et al 1996). Hence, when considering the $dw/dt$ equation for the strain $ay$, the sum in the right-hand side will include contributions from strains $ax$, $ay$ and $by$ but will exclude strain $bx$.

In order to quantify interactions between different strains, it is convenient to introduce an $N\times N$ connectivity matrix $A$, whose entries indicate whether or not two strains are antigenically related. If the antigenic distance between strains is not taken into account, the entries of the matrix $A$ would be zeros if the two variants are immunologically completely distinct, and ones if they are related. Several papers have considered how one can make such a description more realistic by including antigenic distance between different strains, which can be done by using, for instance, the Hamming between two strings representing alleles in the locus of each strain (Adams and Sasaki 2009, Cobey and Pascual 2011, Gog and Grenfell 2002, Gomes et al 2002, Recker and Gupta 2005). Since we are primarily interested in the symmetry properties of the interactions between different strains, we will not consider the effects of antigenic distance on the dynamics.

Before proceeding with the analysis of this system, one can reduce the number of free parameters by scaling time with the average infectious period $(\mu+\sigma)^{-1}$, and we also introduce the basic reproductive ratio $r=\beta/(\mu+\sigma)$ and the ratio of a typical infectious period to a typical host lifetime $e=\mu/(\mu+\sigma)$.
Using the connectivity matrix $A$ and the new parameters, the system (\ref{GFAsys}) can be rewritten as follows
\begin{equation}\label{Sys}
\begin{array}{l}
\displaystyle{\frac{dy_i}{dt}={\Lambda}_i[1-w_i-(1-\gamma)z_i]-y_i,}\\\\
\displaystyle{\frac{dz_i}{dt}={\Lambda}_i(1-z_i)-ez_i,}\\\\
\displaystyle{\frac{dw_i}{dt}=(1-w_i)(A{\bf \Lambda})_i-ew_i,}
\end{array}
\end{equation}
where $\Lambda_i=ry_i$ and ${\bf \Lambda}=(\Lambda_1,\Lambda_2,\ldots,\Lambda_N)^{T}$.

For the particular antigenic system shown in Fig.~\ref{strains}, if one enumerates the strains as follows,
\begin{equation}\label{var4}
\begin{array}{cc}
1&\hspace{0.3cm}ax,\\
2&\hspace{0.3cm}ay,\\
3&\hspace{0.3cm}by,\\
4&\hspace{0.3cm}bx,
\end{array}
\end{equation}
the corresponding connectivity matrix is given by
\begin{equation}\label{Amat}
A=\left(
\begin{array}{cccc}
1&1&0&1\\
1&1&1&0\\
0&1&1&1\\
1&0&1&1
\end{array}
\right).
\end{equation}
The construction of the connectivity matrix can be generalized to an arbitrary number of loci and alleles.

The above system has to be augmented by appropriate initial conditions, which are taken to be
\[
y_i\geq 0,z_i(0)\geq 0,w_i\geq 0.
\]
It is straightforward to show that with these initial conditions, the system (\ref{Sys}) is well-posed in that its solutions remain non-negative for all time.

\section{Symmetry analysis of steady states}

\begin{figure}
\hspace{1cm}
\epsfig{file=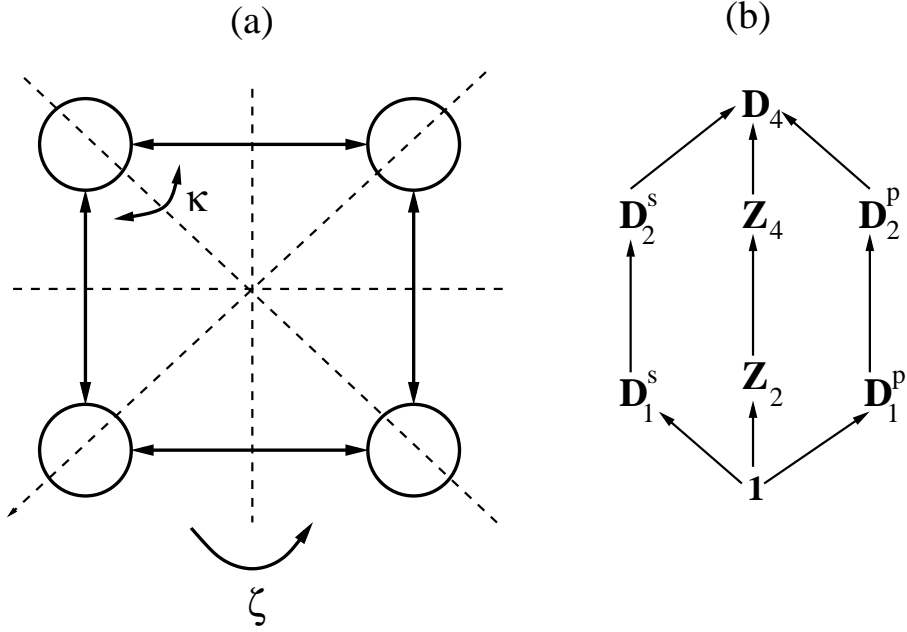,width=12cm}
\caption{(a) Symmetries of the square. (b) Lattice of subgroups of ${\bf D}_4$ symmetry group.}\label{d4sym}
\end{figure}

System (\ref{Sys}) has a large number of biologically realistic steady states. As expected, the trivial steady state $\mathcal{O}=\{y_i=z_i=w_i=0\}$ is unstable when the basic reproductive ratio $r$ exceeds unity. In order to systematically study other steady states and their stability, as well as to illustrate how the methods of equivariant bifurcation theory can be employed to obtain useful insights into stability and dynamics of the system, we concentrate on a specific connectivity matrix $A$ given in (\ref{Amat}) that corresponds to a two locus-two allele system (\ref{var4}).
In this case $N=4$, and the system (\ref{Sys}) is equivariant under the action of a dihedral group ${\bf D}_{4}$, which is an 8-dimensional symmetry group of a square. This group can be written as ${\bf D}_{4}=\{1,\zeta,\zeta^2,\zeta^3,\kappa,\kappa\zeta,\kappa\zeta^2,\kappa\zeta^3\}$, and it is generated by a four-cycle $\zeta$ corresponding to counterclockwise rotation by $\pi/2$, and a flip $\kappa$, whose line of reflection connects diagonally opposite corners of the square, see Fig.~\ref{d4sym}(a).

The group ${\bf D}_{4}$ has eight different subgroups (up to conjugacy): ${\bf 1}$, ${\bf Z}_{4}$, and ${\bf D}_4$, as well as ${\bf D}_{1}^{p}=\{1,\kappa \}$ generated by a reflection across a diagonal, ${\bf D}_{1}^{s}=\{1,\kappa\zeta\}$ generated by a reflection across a vertical, ${\bf D}_{2}^{p}=\{1,\zeta^2,\kappa,\kappa\zeta^2\}$ generated by reflections across both diagonals, and ${\bf D}_{2}^{s}=\{1,\zeta^2,\kappa\zeta,\kappa\zeta^3\}$ generated by the horizontal and vertical reflections. Finally, the group ${\bf Z}_2$ is generated by rotation by $\pi$. The lattice of these subgroups is shown in Fig.~\ref{d4sym}(b). The group ${\bf D}_{4}$ has two other subgroups ${\bf Z}_2(\kappa\zeta^2)=\{1,\kappa\zeta^2\}$ and ${\bf Z}_2(\kappa\zeta^3)=\{1,\kappa\zeta^3\}$, which will be omitted as they are conjugate to ${\bf D}_{1}^{p}$ and ${\bf D}_{1}^{s}$, respectively. There is a certain variation in the literature regarding the notation for subgroups of ${\bf D}_4$, and we are using the convention adopted in Golubitsky and Stewart (2002), c.f. (Buono and Golubitsky 2001, Golubitsky et al 1988).

The group ${\bf D}_4$ has four one-dimensional irreducible representations (F\"assler and Stiefel 1992, Golubitsky and Stewart 1986). Equivariant Hopf Theorem (Golubitsky et al 1988, Golubitsky and Stewart 2002) states that under certain genericity hypotheses, there exists a branch of small-amplitude periodic solutions corresponding to each ${\bf C}$-{\it axial} subgroup $\Gamma\times{\bf S}^1$ acting on the centre subspace of the equilibrium. To find out what type of periodic solution the fully symmetric steady state will actually bifurcate to, we can use the subspaces associated with the above-mentioned one-dimensional irreducible representations to perform an isotypic decomposition of the full phase space (Blyuss 2013, Swift 1988).

The find the fully symmetric steady state (i.e. when all strains are exactly the same)  we can look for it in the form
\[
y_{i}=Y,\hspace{0.5cm}z_i=Z,\hspace{0.5cm}w_i=W,\hspace{0.5cm}i=1,2,3,4.
\]
Substituting this into (\ref{Sys}) gives the system of coupled equations
\begin{equation}\label{FSSS}
\begin{array}{l}
r[1-\gamma W-(1-\gamma)Z]=1,\\\\
rY(1-Z)-eZ=0,\\\\
3rY(1-W)-eW=0.
\end{array}
\end{equation}
The last two equations can be solved to yield
\begin{equation}\label{ZW}
\displaystyle{Z=\frac{rY}{rY+e},\hspace{0.5cm}W=\frac{3rY}{3rY+e}},
\end{equation}
and substituting these expressions into the first equation of (\ref{FSSS}) gives the quadratic equation for $Y$:
\[
3r^2Y^2-re[r(3-2\gamma)-4]Y-(r-1)e^2=0.
\]
This equation can only have a positive root for $r>1$:
\begin{equation}
\displaystyle{Y=\frac{e}{6r}\left[r(3-2\gamma)-4+\sqrt{[r(3-2\gamma)-4]^2+12(r-1)}\right].}
\end{equation}
Hence, the fully symmetric steady state is given by
\begin{equation}
E=(Y,Y,Y,Y,Z,Z,Z,Z,W,W,W,W),
\end{equation}
and it only exists for $r>1$, which, expectedly, is exactly the condition of instability of the trivial steady state.

For the fully symmetric steady state $E$, the Jacobian of linearization takes the block form
\begin{equation}\label{JacE}
J(E)=\left(
\begin{array}{ccc}
{\bf 0}_4 &-\beta(1-\gamma)Y {\bf 1}_4 &-\beta\gamma Y {\bf 1}_4\\
\beta(1-Z) {\bf 1}_4 & -(\mu+\beta Y) {\bf 1}_4& {\bf 0}_4\\
\beta(1-W) A & {\bf 0}_4 & -(\mu+3\beta Y) {\bf 1}_4
\end{array}
\right),
\end{equation}
where ${\bf 0}_4$ and ${\bf 1}_4$ are $4\times 4$ zero and unit matrices, and $A$ is the connectivity matrix (\ref{Amat}). Rather than compute stability eigenvalues directly from this $12\times 12$ matrix, we can use isotopic decomposition of the phase space to block-diagonalize this Jacobian. We note that ${\bf D}_4$ acts to permute indices of different strains, hence our phase space $({\mathbb R}^4)^3$ consists of three copies of the irreducible representations of ${\mathbb R}^4$. Dellnitz and Melbourne (1994) have shown earlier that the sub-spaces
\begin{equation}\label{IsD}
\mathbb{R}\{(1,1,1,1)\},\hspace{0.5cm}\mathbb{R}\{(1,-1,1,-1)\},\hspace{0.5cm}\mathbb{R}\{(1,0,-1,0),(0,1,0,-1)\},
\end{equation}
are ${\bf D}_4$-irreducible and give isotypic components of ${\mathbb R}^4$. Using such decomposition on $(y_1,y_2,y_3,y_4)^{T}\in\mathbb{R}^4$, $(z_1,z_2,z_3,z_4)^{T}\in\mathbb{R}^4$ and $(w_1,w_2,w_3,w_4)^{T}\in\mathbb{R}^4$, the Jacobian (\ref{JacE}) can be block-diagonalized in the following way (Golubitsky and Stewart 1986, Swift 1988):
\begin{equation}\label{JBD}
J^{BD}=BJ(E)B^{-1}=\left(
\begin{array}{cccc}
C+2D & {\bf 0}_3&{\bf 0}_3&{\bf 0}_3\\
{\bf 0}_3&C-2D&{\bf 0}_3&{\bf 0}_3\\
{\bf 0}_3&{\bf 0}_3&C&{\bf 0}_3\\
{\bf 0}_3&{\bf 0}_3&{\bf 0}_3&C
\end{array}
\right),
\end{equation}
where the matrix
\[
B=\left(
\begin{array}{cccccccccccc}
1&1&1&1&0&0&0&0&0&0&0&0\\
0&0&0&0&1&1&1&1&0&0&0&0\\
0&0&0&0&0&0&0&0&1&1&1&1\\
1&0&-1&0&0&0&0&0&0&0&0&0\\
0&0&0&0&1&0&-1&0&0&0&0&0\\
0&0&0&0&0&0&0&0&1&0&-1&0\\
1&-1&1&-1&0&0&0&0&0&0&0&0\\
0&0&0&0&1&-1&1&-1&0&0&0&0\\
0&0&0&0&0&0&0&0&1&-1&1&-1\\
0&1&0&-1&0&0&0&0&0&0&0&0\\
0&0&0&0&0&1&0&-1&0&0&0&0\\
0&0&0&0&0&0&0&0&0&1&0&-1
\end{array}
\right),
\]
is the transformation matrix based on the isotopic decomposition (\ref{IsD}), and
\begin{equation}\label{CD}
C=\left(
\begin{array}{ccc}
0&-\beta(1-\gamma)Y&-\beta\gamma Y\\
\beta(1-Z)&-\beta Y-\mu&0\\
\beta(1-W)&0&-3\beta Y-\mu
\end{array}
\right),\hspace{0.5cm}
D=\left(
\begin{array}{ccc}
0&0&0\\
0&0&0\\
\beta(1-W)&0&0
\end{array}
\right).
\end{equation}
Here, matrix $C$ is associated with self-coupling, and $D$ is associated with nearest-neighbour coupling. Isotypic decomposition of the phase space results in representation of this space as a direct sum of three linear subspaces (Swift 1988)
\[
{\bf R}^{12}=V_{e}\oplus V_{o}\oplus V_{4},
\]
where $V_{e}\simeq{\mathbb R}^3$, called `even' subspace, is the invariant subspace where all strains behave identically the same; $V_{o}\simeq{\mathbb R}^3$, known as `odd' subspace, has each strain being in anti-phase with its neighbours, and in the subspace $V_4$, each strain is in anti-phase with its diagonal neighbour. ${\bf D}_4$-invariance of these subspaces implies that stability changes in the $C+2D$, $C-2D$ and $C$ matrices describe a bifurcation of the fully symmetric steady state $E$ in the even, odd, and $V_4$ subspaces, respectively (Swift 1988). Prior to performing stability analysis, we recall the Routh-Hurwitz criterion, which states that all roots of the equation
\[
\lambda^3+a_1\lambda^2+a_2\lambda+a_3=0,
\]
are contained in the left complex half-plane (i.e. have negative real part), provided the following conditions hold (Murray 2002)
\begin{equation}\label{RHcon}
\begin{array}{l}
a_i>0,\hspace{0.5cm}i=1,2,3,\\\\
a_1a_2>a_3.
\end{array}
\end{equation}
The above cubic equation has a pair of purely imaginary complex conjugate eigenvalues when
\begin{equation}\label{RHhopf}
\begin{array}{l}
a_i>0,\hspace{0.5cm}i=1,2,3,\\\\
a_1a_2=a_3,
\end{array}
\end{equation}
as discussed by Farkas and Simon (1992).\\

\noindent {\bf Theorem 1.} {\it The fully symmetric steady state $E$ is stable when}
\[
K_1>0,\hspace{0.5cm}K_2>0,\hspace{0.5cm}K_3>0,
\]
{\it where}
\[
\begin{array}{l}
\displaystyle{K_1=e^2+3r^2Y^2+[2r^2\gamma(W-1)+r+4re]Y,}\\
\displaystyle{K_2=Yr\gamma(W-4)+3Y(1+r^2\gamma W)+2\gamma e(W-1)-e/r,}\\
\displaystyle{K_3=12rY(e^2+r^2Y^2)+r^2 Ye(1-2\gamma)+22r^2Y^2e+2e^3}.
\end{array}
\]
{\it The steady state $E$ is unstable whenever any of the above conditions are violated; it undergoes a Hopf bifurcation in the odd subspace at} $K=0$, {\it and a steady-state bifurcation at} $K_1=0$ {\it and} $K_2=0$.\\

\noindent The proof of this theorem is given in the Appendix.\\

The implication of the fact that the Hopf bifurcation can only occur in the odd subspace of the phase space (Swift 1988) is that in the system (\ref{Sys}) the fully symmetric state $E$ can only bifurcate to an odd periodic orbit, for which strains $ax$ and $by$ are synchronized and half a period out-of-phase with strains $ay$ and $bx$, i.e. each strain is in anti-phase with its nearest antigenic neighbours.

Besides the origin $\mathcal{O}$ and the fully symmetric equilibrium $E$, the system (\ref{Sys}) possesses 14 more steady states characterized by a different number of non-zero strains ${\bf y}$. There are four distinct steady states with a single non-zero strain $y_{i}$, which all have the isotropy subgroup ${\bf D}_1^p$ or its conjugate. A representative steady state of this kind is
\begin{equation}\label{E1}
E_1=(Y_1,0,0,0,Z_1,0,0,0,W_1,W_1,0,W_1).
\end{equation}
with the other steady states $E_2$, $E_3$ and $E_4$ being related to $E_1$ through elements of a subgroup of rotations ${\bf Z}_4$. The values of $Y_1$, $Z_1$ and $W_1$ are determined by the system of equations
\[
\begin{array}{l}
r[1-\gamma W_1-(1-\gamma)Z_1]=1,\\\\
rY_1(1-Z_1)-eZ_1=0,\\\\
r(1-W_1)Y_1-eW_1=0,
\end{array}
\]
which can be solved to yield
\begin{equation}
Y_{1}=e\frac{r-1}{r},\hspace{0.5cm}Z_1=W_1=\frac{rY}{rY+e}.
\end{equation}
Similarly to the fully symmetric steady state, the steady states with a single non-zero variant are only biologically feasible for $r>1$.\\

\noindent{\bf Theorem 2.} {\it All steady states} $E_1$, $E_2$, $E_3$, $E_4$ {\it with one non-zero strain are unstable.}\\

\noindent The proof of this theorem is given in the Appendix.\\

Before moving to the case of two non-zero strains, it is worth noting that elements of the symmetry group ${\bf D}_4$ representing reflections split into two distinct conjugacy classes: reflections along the diagonals of the square, and reflections along horizontal/vertical axes. These two conjugacy classes are related by an outer automorphism, which can be represented as a rotation through $\pi/4$, which is a half of the minimal rotation in the dihedral group ${\bf D}_{4}$ (Golubitsky et al 1988).

Now we consider the case of two non-zero strains, for which there are exactly six different steady states. The steady states with non-zero strains being nearest neighbours in Fig.~(\ref{strains}), i.e. (1,2), (2,3), (3,4) and (1,4), form one cluster:
\[
\begin{array}{l}
E_{12}=(Y_{2},Y_{2},0,0,Z_{2},Z_{2},0,0,W_{22},W_{22},W_{21},W_{21}),\\
E_{23}=(0,Y_{2},Y_{2},0,0,Z_{2},Z_{2},0,W_{21},W_{22},W_{22},W_{21}),\\
E_{34}=(0,0,Y_{2},Y_{2},0,0,Z_{2},Z_{2},W_{21},W_{21},W_{22},W_{22}),\\
E_{14}=(Y_{2},0,0,Y_{2},Z_{2},0,0,Z_{2},W_{22},W_{21},W_{21},W_{22}),
\end{array}
\]
while the steady states with non-zero strains lying across each other on the diagonals, i.e. (1,3) and (2,4), are in another cluster
\[
\begin{array}{l}
E_{13}=(Y_{3},0,Y_{3},0,Z_{3},0,Z_{3},0,W_{31},W_{32},W_{31},W_{32}),\\
E_{24}=(0,Y_{3},0,Y_{3},0,Z_{3},0,Z_{3},W_{32},W_{31},W_{32},W_{31}),
\end{array}
\]
The difference between these two clusters of steady states is in the above-mentioned conjugacy classes of their isotropy subgroups: the isotropy subgroup of the first cluster belongs to a conjugacy class of reflections along the horizontal/vertical axes, with a centralizer given by ${\bf D}_2^s$, and the isotropy subgroup of the second cluster belongs to a conjugacy class of reflections along the diagonals, with a centralizer given by ${\bf D}_2^p$.

Substituting the general expression for the steady state $E_{12}$ into the system (\ref{Sys}) shows that the values of $Y_2$, $Z_2$, $W_{21}$ and $W_{22}$ are determined by the following system of equations
\[
\begin{array}{l}
r[1-\gamma W_{22}-(1-\gamma)Z_2]=1,\\\\
rY_2(1-Z_2)=eZ_2,\\\\
rY_2(1-W_{21})=eW_{21},\\\\
2rY_2(1-W_{22})=eW_{22}.
\end{array}
\]
The last three equations of this system can be solved in a straightforward way to give
\begin{equation}
Z_{2}=\frac{rY_2}{rY_2+e},\hspace{0.5cm}W_{21}=\frac{rY_2}{rY_2+e},\hspace{0.5cm}W_{22}=\frac{2rY_2}{2rY_2+e},
\end{equation}
and substituting this into the first equation of the system gives the quadratic equation for $Y_2$
\[
2r^2Y_2^2+re[r(\gamma-2)+3]Y-e^2(r-1)=0,
\]
with the solution
\begin{equation}\label{Y2def}
Y_2=\frac{e}{4r}\left[r(2-\gamma)-3+\sqrt{[r(2-\gamma)-3]^2+8(r-1)}\right],
\end{equation}
and this solution is biologically feasible only for $r>1$.

In a very similar way, substituting the expected form of the steady state $E_{13}$ into the system (\ref{Sys}) gives the following system of equations for $Y_{3}$, $Z_3$, $W_{31}$, $W_{32}$
\[
\begin{array}{l}
r[1-\gamma W_{31}-(1-\gamma)Z_3]=1,\\\\
rY_3(1-Z_3)=eZ_3,\\\\
rY_3(1-W_{31})=eW_{31},\\\\
2rY_3(1-W_{32})=eW_{32}.
\end{array}
\]
Once again, we first solve the last three equations to find
\begin{equation}
Z_{3}=\frac{rY_3}{rY_3+e},\hspace{0.5cm}W_{31}=\frac{r3_2}{rY_3+e},\hspace{0.5cm}W_{32}=\frac{2rY_3}{2rY_3+e},
\end{equation}
and substituting them into the first equation of the above systems yields the value of $Y_3$ as
\begin{equation}
\displaystyle{Y_3=e\frac{r-1}{r},}
\end{equation}
and one can note that this steady state is again only biologically feasible when $r>1$.\\

\noindent{\bf Theorem 3.} {\it All steady states} $E_{12}$, $E_{23}$, $E_{34}$, $E_{14}$, {\it are unstable. Steady states } $E_{13}$ {\it and} $E_{24}$, {\it are stable for}\\
\begin{equation}\label{stab_e13}
\displaystyle{r<r_c=\frac{1}{2(1-\gamma)},}
\end{equation}
{\it and unstable otherwise}.\\

\noindent The proof of this theorem is given in the Appendix.\\

For three non-zero variants, we again have four different steady states having an isotropy subgroup ${\bf D}_1^p$ or its conjugate, with a representative steady state being
\[
E_{124}=(Y_{41},Y_{42},0,Y_{42},Z_{41},Z_{42},0,Z_{42},W_{41},W_{42},W_{43},W_{42}),
\]
and the other steady states $E_{123}$, $E_{234}$ and $E_{134}$ being related to $E_{124}$ through elements of a subgroup of rotations ${\bf Z}_4$. Substituting this form of the steady state into the system (\ref{Sys}) shows that the different components of $E_{124}$ satisfy
\[
\begin{array}{l}
r[1-\gamma W_{41}-(1-\gamma)Z_{41}]=1,\\\\
r[1-\gamma W_{42}-(1-\gamma)Z_{41}]=1,\\\\
rY_{41}(1-Z_{41})-eZ_{41}=0,\\\\
rY_{42}(1-Z_{42})-eZ_{42}=0,\\\\
r(Y_{41}+2Y_{42})(1-W_{41})-eW_{41}=0,\\\\
r(Y_{41}+Y_{42})(1-W_{42})-eW_{42}=0,\\\\
2rY_{42}(1-W_{43})-eW_{43}=0.
\end{array}
\]
Solving this system in a manner similar to that for other steady states considered earlier yields
\[
\begin{array}{l}
\displaystyle{Y_{41}=\frac{e[1-r+rY_{42}\left(2+r(Y_{42}-1)\right)]}{r[r(1-\gamma)-1-rY_{42}]},\hspace{0.5cm}Z_{41}=\frac{rY_{41}}{rY_{41}+e},\hspace{0.5cm}Z_{42}=\frac{rY_{42}}{rY_{42}+e},}\\\\
\displaystyle{W_{41}=\frac{r(Y_{41}+2Y_{42})}{r(Y_{41}+2Y_{42})+e},\hspace{0.5cm}W_{42}=\frac{r(Y_{41}+Y_{42})}{r(Y_{41}+Y_{42})+e},\hspace{0.5cm}
W_{43}=\frac{2rY_{42}}{2rY_{42}+e}},
\end{array}
\]
and $Y_{42}$ is a positive root of the quartic equation
\[
\begin{array}{l}
\displaystyle{r^2 z^4-r(r-2)z^3+[r^2(4\gamma-1)(1-\gamma)+1-2\gamma r]z^2+\gamma(r-\gamma)(\gamma-1)}\\\\
\displaystyle{+\left[1+r^2+8r\gamma-2(r+\gamma)-r\gamma\left(6\gamma(1-r)+r(5+2\gamma^2)\right)\right]z=0.}
\end{array}
\]
It does not prove possible to find a closed form expression for the eigenvalues of linearization near $E_{124}$, hence these eigenvalues have to be computed numerically. For all biologically realistic values of parameters we have studied, one of these eigenvalues is always positive, suggesting that a steady state $E_{124}$ (and also $E_{123}$, $E_{234}$, $E_{134}$) is unstable.

\begin{figure}
\hspace{0.5cm}
\epsfig{file=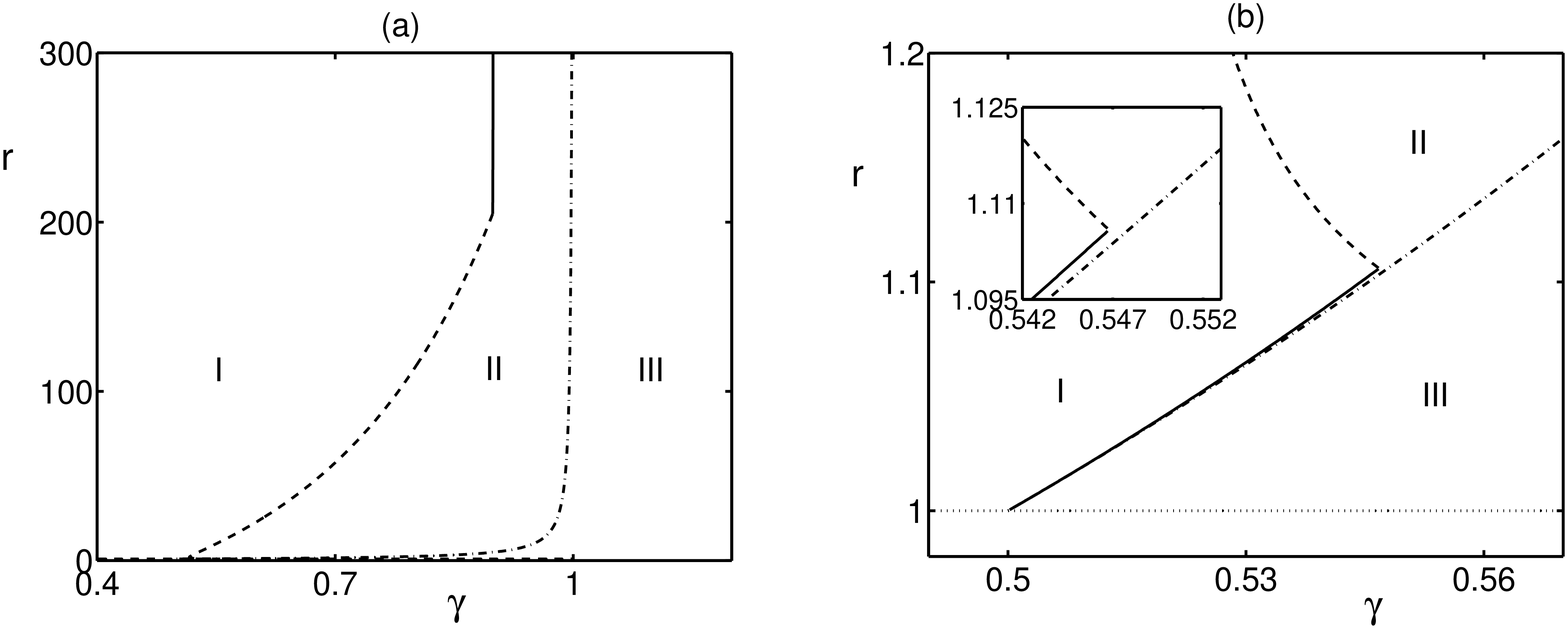,width=15cm}
\caption{Bifurcation diagram for the four-strain system (\ref{Sys}). Parameter values are: $\mu=0.02$, $\sigma=10$. In region I only the fully symmetric steady state $E$ is stable, in region III only the steady states $E_{13}$ and $E_{24}$ are stable, and in the region II all the steady states are unstable. Solid line denotes the boundary of a steady-state bifurcation of the steady state $E$, dashed line is the boundary of Hopf bifurcation of the steady state $E$, dash-dotted line is stability boundary of the steady states $E_{13}$ and $E_{24}$.}\label{BifDia}
\end{figure}

Figure~\ref{BifDia} shows the bifurcation diagram for different steady states depending on the disease transmission rate $\beta$ and the cross-immunity $\gamma$. If $r\leq 1$, the only biologically feasible steady state is the disease-free equilibrium $\mathcal{O}$, and it is stable. When $r>1$, the other steady states with different numbers of non-zero strains are also biologically feasible. For sufficiently small values of cross-immunity $\gamma$, the fully symmetric steady state $E$ is the only stable steady state, and as $\gamma$ increases, this steady state loses its stability either via Hopf bifurcation or via a steady-state bifurcation. When the fully symmetric steady state $E$ undergoes Hopf bifurcation, it gives rise to a stable anti-phase periodic orbit, however as $\gamma$ is increased, this periodic orbit disappears via a global bifurcation upon collision with two steady states $E_{13}$ and $E_{24}$; such behaviour has been observed by Dawes and Gog (2002) who performed a very detailed bifurcation analysis of the case $r\approx 1$. Figure~\ref{BifDia} also shows that although a large number of different non-trivial steady states may exist for $r>1$, when the cross-immunity between strains $\gamma$ is close to one, this will make it impossible for the immunologically closest strains to simultaneously survive, thus resulting in the fact that the only stable steady states in this regime are "edge" equilibria $E_{13}$ and $E_{24}$ with antigenically unrelated strains present (Dawes and Gog 2002).

\section{Dynamical behaviour of the model}

In the previous section we studied stability of different steady states of the system (\ref{Sys}) and found conditions under which a fully symmetric steady state $E$ can undergo Hopf bifurcation, giving rise to a stable anti-phase periodic solution. Now we look at the evolution of this solution and its symmetries under changes in system parameters. For convenience, we fix all parameters except for the cross-immunity $\gamma$, which is taken to be a control parameter.

\begin{figure}
\hspace{0.5cm}
\epsfig{file=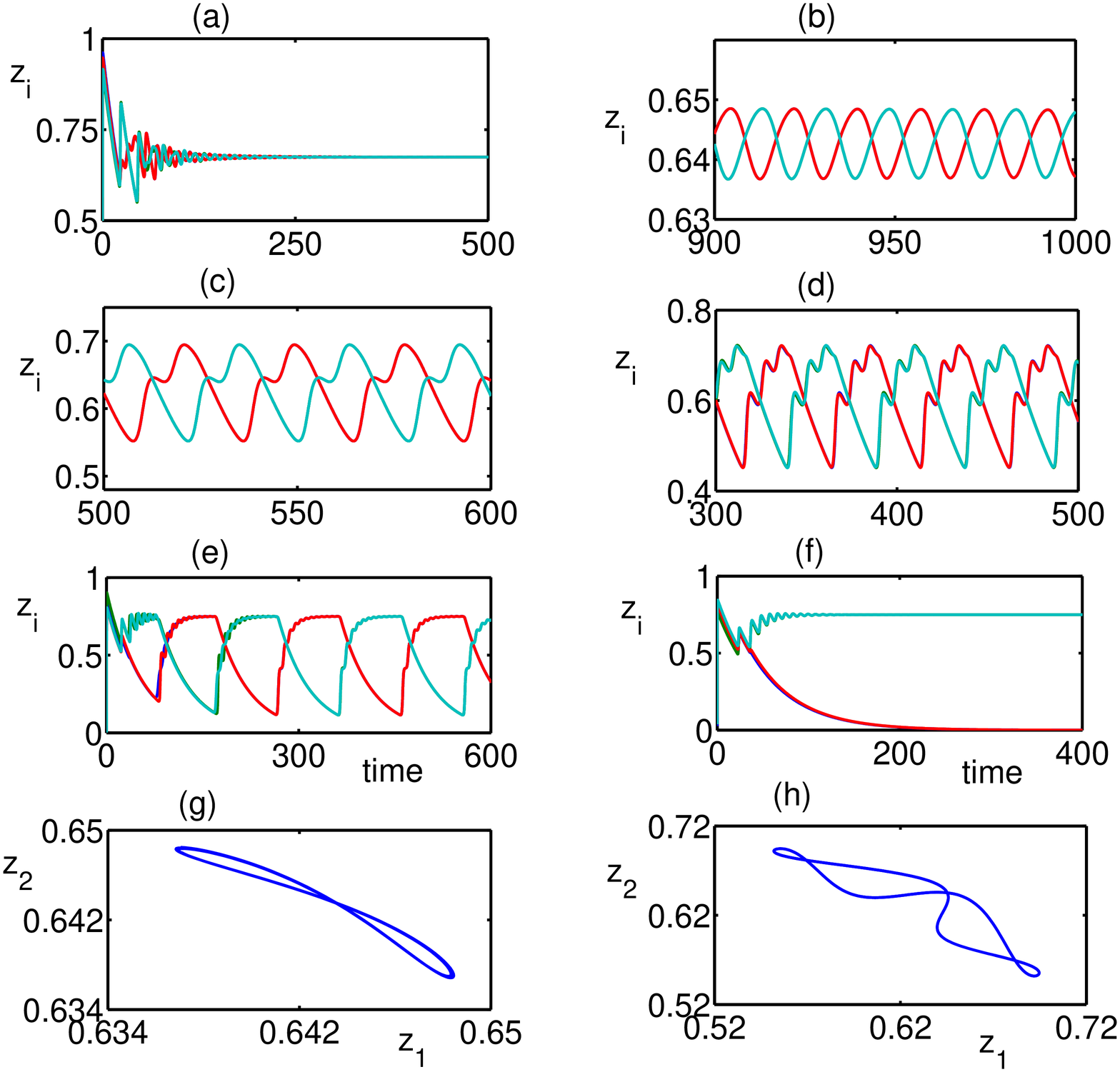,width=15cm}
\caption{Temporal dynamics of the system (\ref{Sys}) with ${\bf D}_4$ symmetry. Parameter values are $\beta=40$, $\sigma=10$, $\mu=0.02$. (a) Stable fully symmetric equilibrium ($\gamma=0.4$). (b-e) Anti-phase periodic solution, $\gamma=0.53$, 0.58, 0.65, 0.8. (f) Stable steady state $E_{13}$ ($\gamma=0.9$). Colours represent different strains: $ax$ (cyan), $ay$ (red). (g) Phase plane for $\gamma=0.53$. (h) Phase plane for $\gamma=0.58$.}\label{dyn_fig}
\end{figure}

The results of numerical simulations are presented in Fig.~\ref{dyn_fig}. When $\gamma$ is sufficiently small, the fully symmetric steady state is stable, as shown in Fig.~\ref{dyn_fig}(a). As $\gamma$ crosses the threshold of Hopf bifurcation as determined by {\bf Theorem 1}, the fully symmetric steady state loses stability, giving rise to an 'odd' periodic solution illustrated in Fig.~\ref{dyn_fig}(b), where strains 1 and 3 are oscillating in complete synchrony and exactly half a period out of phase with strains 2 and 4 which also oscillate in synchrony. Figures~\ref{dyn_fig}(c)-(e) show that for higher values of $\gamma$, the periodic solution remains stable and retains its symmetry but changes the temporal profile.  For very large values of $\gamma$, this periodic orbit becomes unstable, and the system tends to a steady state $E_{13}$ with ${\bf D}_{2}^{p}$ isotropy subgroup, which is stable in the light of {\bf Theorem 3}. In this case, we conclude that the cross-immunity between any two strains which are immunologically closest to each other is so strong that it actually leads to elimination of one of these strains, thus creating a situation where two strains that are most immunologically distant survive, and the other two strains are eradicated. It is worth mentioning that due to the symmetry between the strains, there is no inherent preference for survival of the $(ax,by)$ or $(ay,bx)$ pair of strains.

To classify the symmetry of other possible types of periodic solutions, it is convenient to refer to the $H/K$ Theorem, which uses information about individual spatial and spatio-temporal symmetries of periodic solutions (Buono and Golubitsky 2001, Golubitsky and Stewart 2002). To use this method, we note that due to ${\bf D}_4$-equivariance of the system (\ref{Sys}) and uniqueness of its solutions, it follows that for any $T$-periodic solution $x(t)$ and any element $\gamma\in\Gamma$ of the group, one can write
\[
\gamma x(t)=x(t-\theta),
\]
for some phase shift $\theta\in{\bf S}^{1}\equiv\mathbb{R}/\mathbb{Z}\equiv[0,T)$. The pair $(\gamma,\theta)$ is called a {\it spatio-temporal symmetry} of the solution $x(t)$, and the collection of all spatio-temporal symmetries of $x(t)$ forms a subgroup $\Delta\subset\Gamma\times{\bf S}^{1}$. One can identify $\Delta$ with a pair of subgroups, $H$ and $K$, such that $K\subset H\subset\Gamma$. We also define
\[
\begin{array}{l}
H\hspace{0.3cm}=\hspace{0.3cm}\left\{\gamma\in\Gamma:\gamma\{x(t)\}=\{x(t)\}\right\}\hspace{0.5cm}\mbox{spatio-temporal symmetries},\\
K\hspace{0.3cm}=\hspace{0.3cm}\left\{\gamma\in\Gamma:\gamma x(t)=x(t)\hspace{0.3cm}\forall t\right\}\hspace{0.6cm}\mbox{spatial symmetries.}
\end{array}
\]
\begin{figure}
\hspace{0.5cm}
\epsfig{file=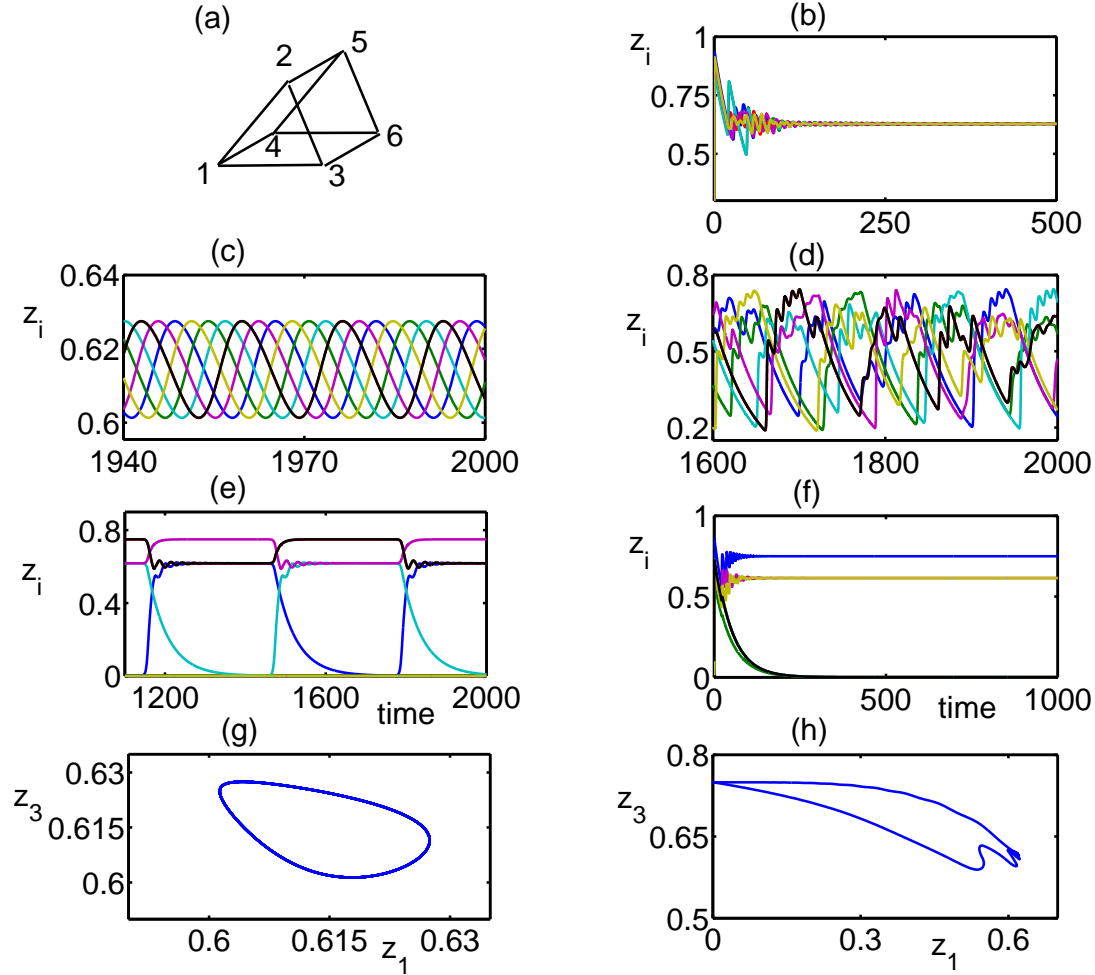,width=15cm}
\caption{(a) Topology of the strain space of the system (\ref{GFA}) with ${\bf S}_3\times{\bf S}_2$ symmetry. (b)-(f) Temporal dynamics of the system (\ref{Sys}) with ${\bf S}_3\times{\bf S}_2$ symmetry. Parameter values are $\beta=40$, $\sigma=10$, $\mu=0.02$. (b) Stable fully symmetric equilibrium ($\gamma=0.5$). (c) Discrete travelling wave, spatio-temporal symmetry $(H,K)=({\bf Z}_6,{\bf 1})$ ($\gamma=0.54$). (d) Chaos ($\gamma=0.75$). (e) Periodic solution with spatio-temporal symmetry $(H,K)=({\bf S}_3\times{\bf S}_2,\sigma_{v})$ ($\gamma=0.9$). (f) Stable stady state ($\gamma=0.92$) with the symmetry $\sigma_{v}$. Colours represent strains 1 to 6. (g) Phase plane for $\gamma=0.54$. (h) Phase plane for $\gamma=0.9$.}\label{GFA6_fig}
\end{figure}
Here, $K$ consists of the symmetries that fix $x(t)$ at each point in time, while $H$ consists of the symmetries that fix the entire trajectory. Under some generic assumptions on $H$ and $K$, the $H/K$ Theorem states that periodic states have spatio-temporal symmetry group pairs $(H,K)$ only if $H/K$ is cyclic, and $K$ is an isotropy subgroup (Buono and Golubitsky 2001, Golubitsky and Stewart 2002). The $H/K$ Theorem was originally derived in the context of equivariant dynamical systems by Buono and Golubitsky (2001), and it has subsequently been used to classify various types of periodic behaviours in systems with symmetry that arise in a number of contexts, from speciation (Stewart 2003) to animal gaits (Pinto and Golubitsky 2006) and vestibular system of vertebrates (Golubitsky et al 2007).

From epidemiological perspective, the spectrum of behaviours that can be exhibited in the case of ${\bf D}_4$ symmetry is quite limited, as it only includes a fully symmetric steady state, a steady state with two non-zero strains, and an anti-phase periodic orbit having a spatio-temporal symmetry with spatio-temporal symmetry $(H,K)=({\bf D}_4,{\bf D}_{2}^{p})$. In order to explore other possible dynamical scenarios, we extend the strain space by assuming that the system (\ref{Sys}) has three alleles in the first locus and two alleles in the second locus. This gives the ${\bf S}_3\times{\bf S}_2$ symmetry group, which is isomorphic to a group ${\bf D_{3h}}$ - dihedral symmetry group of a triangular prism. The results of numerical simulations for such system of strains are shown in Fig.~\ref{GFA6_fig}. For sufficiently small value of $\gamma$, the system again supports a stable fully symmetric steady state in a manner similar to the case of ${\bf D}_4$ symmetry. However, when $\gamma$ exceeds certain threshold, this steady state undergoes Hopf bifurcation, giving rise to a periodic solution, which is a discrete travelling wave with the symmetry $(H,K)=({\bf Z}_6,{\bf 1})$, as shown in Fig.~\ref{GFA6_fig}(c). 
In this dynamical regime all variants appear sequentially one after another with one sixth of a period difference between two neighbouring variants.
From the perspective of equvariant bifurcation theory, this solution is generic since the group ${\bf Z}_{n}$ is always one of the subgroups of the ${\bf D}_n$ group for the ring coupling,  or the ${\bf S}_n$ group for an all-to-all coupling, and its existence has already been extensively studied (Aronson et al 1991, Golubitsky and Stewart 1986, Golubitsky et al 1988). From the epidemiological point of view, this is an extremely important observation that effectively such solution, which represents sequential appearance of antigenically related strains of infection, owes its existence not so much to the individual dynamics of the strains, but rather to the particular symmetric nature of cross-reactive interactions between them.

As the value of $\gamma$ increases, the discrete travelling wave transforms into a quasi-periodic solution, and then to a chaotic solution, where different strains appear in no particular order, and the temporal dynamics of each of them is chaotic, as illustrated in Fig.~\ref{GFA6_fig}(d). For higher values of $\gamma$, the dynamics becomes periodic again, albeit with a different type of spatio-temporal symmetry, given by $(H,K)=({\bf S}_3\times{\bf S}_2,\sigma_{v})$, where $\sigma_v$ is a reflection symmetry with respect to a plane going through the edges 2 and 5, as well as mid-points of the sides 1-3 and 4-6. As $\gamma$ increases further still, the system tends to a stable steady state having the symmetry $\sigma_v$. This steady state is similar to the case of ${\bf D}_4$ symmetry considered earlier in that it contains three non-zero strains, with maximal possible antigenic distance between them.

\section{Other  models of multi-strain dynamics}

The approach developed in the previous section is sufficiently generic and can be applied to the analysis of a variety of different models for multi-strain diseases, where the existence of a degree of cross-protection (or cross-enhancement) between antigenically distinct strains results in a certain symmetry of strain interactions, which then translates into different types of periodic dynamics. Epidemiological data and mathematical models suggest that such systems may exhibit a wide range of behaviours, from no strain structure (NSS), which represents a system approaching a stable steady state, through the discrete or cyclic strain structure (CSS), where the systems demonstrates single strain dominance and sequential trawling through the whole antigenic repertoire, to a chaotic strain structure. Various aspects of the overlapping antigenic repertoires have already been investigated in a number of models, but so far the effects of symmetry in such systems have remained largely unexplored.

As an illustration, we now use symmetry perspective to analyse simulation results in two different multi-strain models. In the first model, analysed by Calvez et al (2005), each strain is characterized by a combination of alleles at immunologically important loci, and the strength of cross-immunity between different strains increases with the number of alleles they share. After some rescaling, the model for such a system can be written in the form
\begin{equation}\label{GFA}
\begin{array}{l}
\displaystyle{\frac{dv_i}{d\tau}=1-(1+y_i)v_i,}\\\\
\displaystyle{\frac{dx_i}{d\tau}=1-\left(1+\sum_{j\sim i}y_j\right)x_i,}\\\\
\displaystyle{\varepsilon_i\frac{dy_i}{d\tau}=[(1-\Gamma_i)v_i+\Gamma_i x_i-r_i]y_i,}
\end{array}
\end{equation}
where $v_i$ is the fraction of individuals who have never been infected with the strain $i$, $x_i$ is the faction of individuals who 
have never been infected with any strain sufficiently close to strain $i$ including strain $i$ itself, $y_i$ is the rescaled fraction of individuals currently infectious with strain $i$, $\varepsilon_i=\mu/\beta_i$ and $r_i=(\mu+\sigma_i)/\beta_i$, $1/\mu$ is the host life expectancy, $1/\sigma$ is an average period of infectiousness, $\beta$ is the transmission rate. Assuming the probability of cross-protection between strains $i$ and $j$ to be $\gamma_{ij}$ (i.e. infection with strain $j$ reduces the probability that the host will be infected by strain $i$ is $\gamma_{ij}$), the force of infection is taken as
\begin{equation}
\Gamma_i=\left(\sum_{j\sim i,j\neq i}\gamma_{ij}y_{j}\right){\Bigg /}\left(\sum_{j\sim i,j\neq i}y_{j}\right).
\end{equation}

\begin{figure}
\hspace{1cm}
\epsfig{file=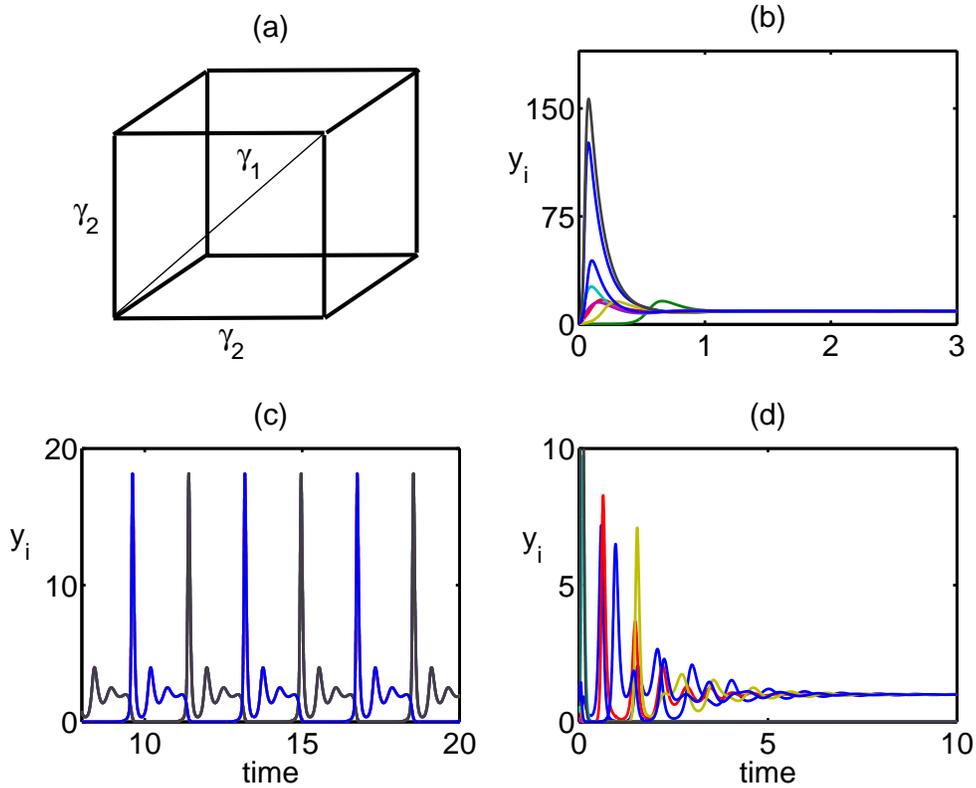,width=15cm}
\caption{(a) Topology of the strain space of the system (\ref{GFA}) with symmetry of cube. (b)-(d). Temporal dynamics of the system (\ref{GFA}) with parameter values $\epsilon=5\cdot 10^{-3}$, $\gamma_1=0.4$, $\gamma_2=0.8$. (b) Stable fully symmetric equilibrium ($r=0.05$). (c) Anti-phase periodic solution with spatio-temporal symmetry $(H,K)=({\bf O},{\bf D}_4)$ ($r=0.25$). (d) ${\bf D}_4$-symmetric stable steady state ($r=0.4$). Colours represent strains 1 to 8.}\label{clus_fig}
\end{figure}

When this system is considered with three loci and two alleles at each locus, this results in an eight-dimensional strain space, as illustrated in Fig.~\ref{clus_fig}(a). Analysis of possible dynamics for such a strain space suggests that "{\it ...it is already not so clear why in the eight-strain system the cluster structure of second type (two clusters of four strains) appears}" (Calvez et al 2005), which is the solution shown in Fig.~\ref{clus_fig}(b)-(c). The authors found this tetrahedral solution unexpected, and indeed stated that "{\it This second type of clustering can hardly be expected a priori}" (Calvez et al 2005). At the same time, when considered from the equivariant bifurcation theory perspective, system (\ref{GFA}) has the octahedral symmetry ${\bf O}$, and therefore, has three maximal isotropy subgroups: the dihedral group ${\bf D_4}$, the permutation group ${\bf S}_3$, and a reflection group ${\bf Z}_2^r\oplus {\bf Z}_2^t$ (Jiang et al 2003, Melbourne 1986). Hence, the bifurcation of a fully symmetric steady state into a tetrahedral periodic solution with ${\bf D}_4$ symmetry should be naturally expected as a result of an equivariant Hopf theorem and an underlying symmetric structure of the antigenic space (Fiedler 1988, Jiang et al 2003).
This example highlights the importance of including symmetry properties of multi-strain epidemic models into consideration of possible steady states and periodic orbits, as it provides a systematic approach to understanding what types of periodic solutions should be expected in the system from a symmetry perspective.

As another example, we consider a model for the population dynamics of dengue fever, which is characterized by an infection with one of four serotypes co-circulating in population. One of the main current theories explaining the observed dynamics of dengue fever is that of antibody-dependent enhancement (ADE), whereby cross-reactive antibodies elicited by a previously encountered serotype bind to the newly infecting heterologous serotype, but fail to neutralize it. This leads to the development of dengue haemorrhagic fever (DHF) and dengue shock syndrome (DSS), characterized by up to 20\% mortality rate (Gubler 2002, Halstead 2007).

In order to explain the observed temporal patterns of disease dynamics, Recker et al (2009) have proposed a model, which assumes that a recovery from an infection with any one serotype is taken to provide permanent immunity against that particular serotype, but it can lead to an enhancement of other serotypes upon secondary infection after which individuals acquire complete immunity against all four serotypes.
In this model the population is divided into the following classes: $s$ denotes the fraction of the population that has not yet been infected with any of the serotypes and is thus totally susceptible; $y_i$ is the proportion infectious with a primary infection with serotype $i$, $r_i$ is the proportion recovered from primary infection with serotype $i$; $y_{ij}$ is the proportion infectious with serotype $j$, having already recovered from infection with serotype $i$; and, finally, $r$ is the proportion of completely
immune (those who have recovered after being exposed to two serotypes). The model equations are given as follows
\begin{equation}\label{reck_eq}
\begin{array}{l}
\displaystyle{\frac{ds}{dt}=\mu-s\sum_{k=1}^{4}\lambda_k-\mu s,}\\\\
\displaystyle{\frac{dy_i}{dt}=s\lambda_i-(\sigma+\mu)y_i,}\\\\
\displaystyle{\frac{dr_i}{dt}=\sigma y_i-r_i\left(\mu+\sum_{j\neq i}\gamma_{ij}\lambda_j\right),}\\\\
\displaystyle{\frac{dy_{ij}}{dt}=r_i\gamma_{ij}\lambda_j-(\sigma+\mu)y_{ij},\hspace{0.3cm}i\neq j,}\\\\
\displaystyle{\frac{dr}{dt}=\sigma\sum_{i=1}^{4}\sum_{j\neq i}y_{ij}-\mu r,}
\end{array}
\end{equation}
where $1/\mu$ is the average host life expectancy, and $1/\sigma$ is the average duration of infectiousness. The force of infection with a serotype $i$, $\lambda_i$ is given by
\[
\lambda_{i}=\beta_i\left(y_i+\sum_{j\neq i}\phi_{ji}y_{ji}\right),
\]
where $\beta_i$ is the transmission coefficient of serotype $i$, and the ADE is represented by two distinct parameters: the enhancement of susceptibility to secondary infections, $\gamma_{ij}\geq 1$, and the enhancement of transmissibility during secondary infection, $\phi_{ij}\geq 1$. Although in this case the antigenic space again consists of four distinct serotypes, but unlike earlier examples of dihedral symmetry the system now has an ${\bf S}_4$ symmetry of four nodes with an all-to-all coupling. For simplicity, it is assumed that all serotypes enhance each other in identical way, i.e. $\gamma_{ij}=\gamma$, and also transmissibility is enhanced in the same way, implying that $\phi_{ij}=\phi$. Hence, we fix all other parameters, and vary $\gamma$ and $\phi$ to explore possible dynamical regimes.

\begin{figure}
\hspace{0.5cm}
\epsfig{file=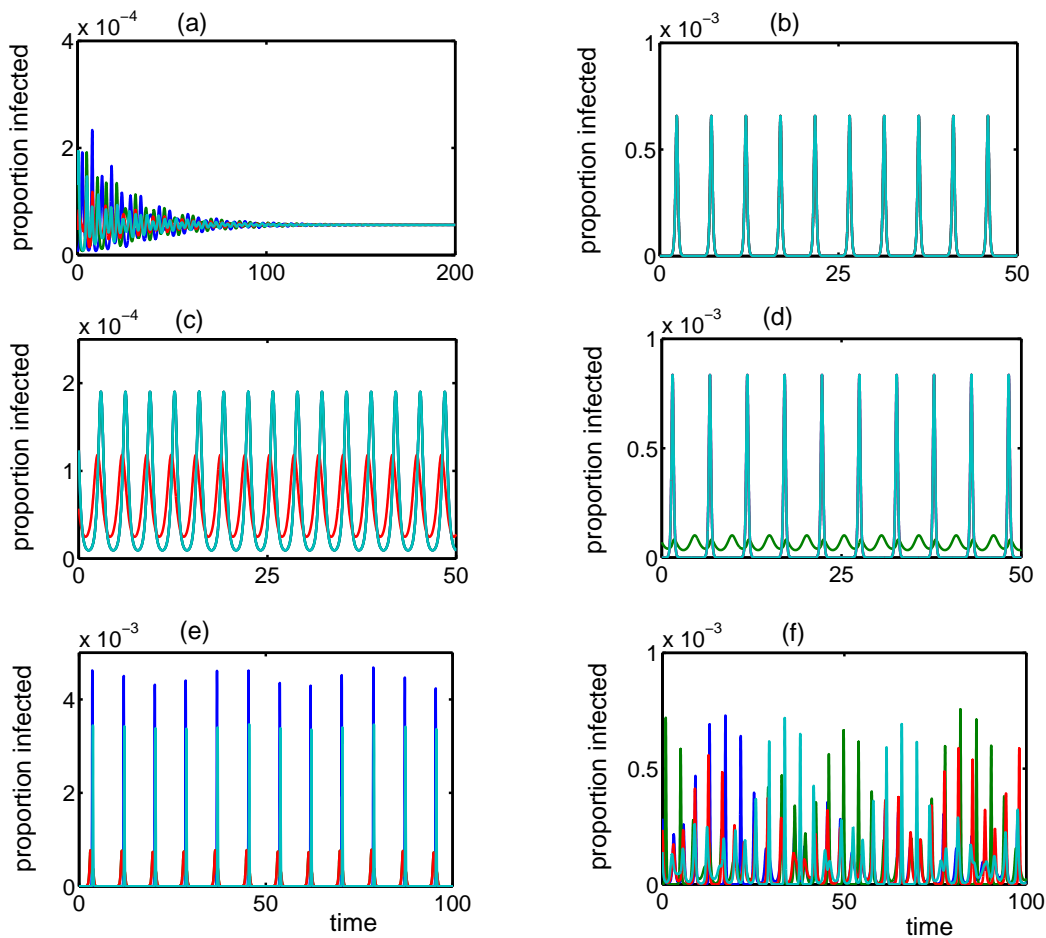,width=15cm}
\caption{Temporal dynamics of the system (\ref{reck_eq}) with ${\bf S}_4$ symmetry. Parameter values are $\beta=200$, $\sigma=100$, $\mu=0.02$. (a) Stable fully symmetric equilibrium ($\gamma=1$, $\phi=1$). (b) Fully symmetric periodic solution, $\gamma=1$, $\phi=2.4$. (c) Periodic solution with spatio-temporal symmetry
$H/K=({\bf S}_4,{\bf S}_3\times {\bf S}_1)$ ($\gamma=1$, $\phi=2.7$). (d) Periodic solution with spatio-temporal symmetry $H/K=({\bf S}_4,{\bf D}_2^{p})$ ($\gamma=2.5$, $\phi=1$). (e) Quasi-periodic solution ($\gamma=2$, $\phi=2$). (f) Chaotic solution ($\gamma=2.5$, $\phi=2.5$). Colours correspond to serotypes 1 (red), 2 (cyan), 3 (green) and 4 (blue).}\label{den_fig}
\end{figure}

Figure~\ref{den_fig} illustrates different types of behaviour that can be exhibited by the system (\ref{reck_eq}) as the enhancement of susceptibility $\gamma$ and enhancement of transmissibility $\phi$ are varied. In the case when both $\gamma$ and $\phi$ are sufficiently small (equal to or just above 1), the system approaches a stable fully symmetric steady state shown in Fig.~\ref{den_fig}(a). As the enhancement of transmissibility $\phi$ increases, the fully symmetric steady state loses stability via a Hopf bifurcation, giving rise to a fully symmetric periodic orbit, as illustrated in Fig.~\ref{den_fig}(b). Depending on the values of $\gamma$ and $\phi$, it is possible to observe other types of periodic solutions: a solution where three serotypes have identical dynamics, and the fourth serotype has a different dynamics (see Fig.~\ref{den_fig}(c)), and a solution with the symmetry of reflections across diagonals shown in Fig.~\ref{den_fig}(d), where antigenically distinct strains have the same behaviour. For higher values of $\gamma$ and $\phi$, the dynamics becomes quasi-periodic and eventually chaotic.

\section{Discussion}

In this paper we have shown how one can use the techniques of equivariant bifurcation theory to systematically approach the analysis of stability of steady states and classification of different periodic solutions in multi-strain epidemic models. Once the underlying symmetry of the system has been established, the steady states can be grouped together using conjugacy classes of the corresponding isotropy subgroups, which significantly reduces computational effort associated with studying their stability. Moreover, isotypic decomposition of the phase space based on irreducible representations of the symmetry group provides a convenient way of identifying the specific symmetry of a periodic solution emerging from a Hopf bifurcation of the fully symmetric equilibrium. The $H/K$ Theorem provides an account of possible types of spatial and temporal symmetries that can be exhibited by periodic solutions, and hence is very useful for systematic classification of observed periodic behaviours.

An important question is to what degree real multi-strain diseases can be efficiently described by mathematical models with symmetry, bearing in mind that in reality systems of antigenic strains may not always fully preserve the assumed symmetry. There are several observations suggesting that the results of analysis of symmetric models are still applicable for understanding the dynamics of real multi-strain infections. The first of these comes from the fact that many features of the model solutions, such as single-strain dominance and sequential appearance of antigenically related strains in a manner similar to the discrete travelling wave solution discussed earlier, are also observed in epidemiological data (Gupta et al 1998, Minaev and Ferguson 2009, Recker et al 2009, Recker et al 2007). Another reason why the conclusions drawn from symmetric models may still hold stems from an argument based on normal hyperbolicity, which is a generic property in such models, suggesting that the main phenomena associated with symmetric models survive under perturbations, including symmetry-breaking perturbations. The discussion of this issue in the context of modelling sympatric speciation using symmetric models can be found in Golubitsky and Stewart (2002). Andreasen et al (1997) have discussed the situation when the basic reproductive ratios of different strains may vary, showing that in this case the endemic equilibrium persists and can still give rise to stable periodic oscillations through a Hopf bifurcation. Similar issue was discussed by Dawes and Gog (2002) who also noted that despite the possibility of oscillations in multi-strain models, quite often the period of such oscillations is comparable to the host lifetime and hence is much longer than the periodicity of real epidemic outbreaks. One possibility how this limitation may be overcome is when there is a sufficiently large number of co-circulating strains, so that the combinations of some of them rising or falling would result in a rapid turnover of the dominant strain, as has been shown in Gupta et al (1998). Reaching a definitive conclusion regarding the validity of symmetric or almost-symmetric multi-strain models requires a precise measurement of population-level transmission rates individual strains, as well as degrees of immunological cross-protection or cross-enhancement, and despite major advances in viral genotyping and infectious disease surveillance, this still remains a challenge.

A really important methodological advantage of the approach presented in this paper is its genericity in a sense that the analysis of stability and periodic dynamics relies on the symmetries in immunological interactions between strains, rather than any specific information regarding their individual dynamics as prescribed by a disease under consideration. The fact that some of the fundamental dynamical features in the behaviour of multi-strain diseases appear to be universal suggests a possibility to make significant inroads in the understanding generic types of dynamics using the analysis of some recurring motifs of strain interactions with relatively simple topology. In the model analysed in this paper, we were primarily concerned with symmetric properties of the matrix of antigenic connectivity and assumed that the strength of immunological cross-reactivity is the same for all strains. One can make the model more realistic by explicitly including the antigenic distance between strains in manner similar to the Hamming distance (Adams and Sasaki 2009, Calvez et al 2005, Recker and Gupta 2005), which would not alter the topology of the network of antigenic variants but introduce different weights for connections between different strains in the network. Another possibility is to consider the effects of time delay in latency or temporary immunity (Arino and van den Driessche 2006, Blyuss and Kyrychko 2010, Lloyd 2001), which although known to play an important role in disease dynamics, have so far not been studied in the context of multi-strain diseases.

\section*{Acknowledgements}
The author would like to thank Jon Dawes for useful discussions and anonymous referees for their helpful comments and suggestions.

\section*{Appendix}

This Appendix contains detailed proofs of {\bf Theorems 1-3}.\\

\noindent{\bf Proof of Theorem 1.} Stability of the fully symmetric steady state $E$ changes when one of the eigenvalues of the Jacobian (\ref{JBD}) goes through zero along the real axis or a pair of complex conjugate eigenvalues crosses the imaginary axis. Due to the block-diagonal form of the Jacobian it suffice to consider separately possible bifurcations in the matrices $C$, $C\pm 2D$.

For the matrix $C$ given in (\ref{CD}), the characteristic equation takes the form
\[
\lambda^3+a_1\lambda^2+a_2\lambda+a_3=0,
\]
with
\[
\begin{array}{l}
a_1=4rY+2e>0,\\\\
\displaystyle{a_2=\frac{2r^2Y^2e(11e+12rY)+rYe[r^2Y(3-2\gamma)+e(r+8e)]+e^4+9r^4Y^4}{(e+rY)(e+3rY)}>0,}\\\\
\displaystyle{a_3=\frac{r^2 Ye[r^2 Y^2(9-8\gamma)+rYe(6-4\gamma)+e^2]}{(e+rY)(r+3rY)}>0.}
\end{array}
\]
In this case $a_{1,2,3}>0$, and also
\[
\begin{array}{l}
\displaystyle{a_1a_2-a_3=\frac{(4rY+2e)[2r^2Y^2e(11e+12rY)+rYe[r^2Y(3-2\gamma)+e(r+8e)]]}{(e+rY)(e+3rY)}}\\\\
\displaystyle{+\frac{e^4+9r^4Y^4-r^2 Ye[r^2 Y^2(9-8\gamma)+rYe(6-4\gamma)+e^2]}{(e+rY)(e+3rY)}}\\\\
\displaystyle{=\frac{(e+rY)(e+3rY)[12rY(rY+e)+r^2Ye(1-2\gamma)+22r^2Y^2e+2e^3]}{(e+rY)(e+3rY)}=}\\\\
\displaystyle{=12rY(rY+e)+r^2Ye+22r^2Y^2e+2e^3>0,}
\end{array}
\]
which, according to the Routh-Hurwitz conditions (\ref{RHcon}), implies that all eigenvalues of the matrix $C$ are contained in the left complex half-plane for any values of system parameters. This means that the steady state $E$ is stable in the $V_4$ subspace.

Similarly, for the matrix $C+2D$ we have the coefficients of the characteristic equation as
\[
\begin{array}{l}
a_1=4rY+2e>0,\\\\
\displaystyle{a_2=\frac{e^2[e^2+r^2Y(1+2\gamma)]+rY[8e^3+9r^3Y^3+rYe(22e+24rY+3r^2Ye)]}{(e+rY)(e+3rY)}>0,}\\\\
\displaystyle{a_3=\frac{r^2 Ye[3r^2Y^2(3-2\gamma)+\mu^2(1+2\gamma)+6erY]}{(e+rY)(e+3rY)}>0,}
\end{array}
\]
and also
\[
\begin{array}{l}
\displaystyle{a_1a_2-a_3=\frac{(4rY+2e)e\left[e[e^2+r^2Y(1+2\gamma)]+r^2Y^2(22e+24rY+3r^2Ye)\right]}{(e+rY)(e+3rY)}}\\\\
\displaystyle{+\frac{rY[8e^3+9r^3Y^3]-r^2 Ye[3r^2Y^2(3-2\gamma)+\mu^2(1+2\gamma)+6erY]}{(e+rY)(e+3rY)}}\\\\
\displaystyle{=\frac{(e+rY)(e+3rY)[12rY(r^2Y^2+e^2)+r^2Ye(1+2\gamma)+2e(11r^2Y^2+e^2)]}{(e+rY)(e+3rY)}}\\\\
=12rY(r^2Y^2+e^2)+r^2Ye(1+2\gamma)+2e(11r^2Y^2+e^2)>0.
\end{array}
\]
Once again, using Routh-Hourwitz conditions (\ref{RHcon}) we conclude that the eigenvalues of the matrix $C+2D$ are always contained in the left complex half-plane, implying stability of the steady state $E$ in the even subspace.

Finally, for the matrix $C-2D$, the coefficients of the characteristic equation are
\[
\begin{array}{l}
a_1=4rY+2e>0,\\\\
\displaystyle{a_2=e^2+3r^2Y^2+[2r^2\gamma(W-1)+r+4re]Y,}\\\\
\displaystyle{a_3=Yr\gamma(W-4)+3Y(1+r^2\gamma W)+2\gamma e(W-1)-e/r,}
\end{array}
\]
Substituting the value of $W$ from (\ref{ZW}) and computing $a_1a_2-a_3$ gives
\[
a_1a_2-a_3=12r Y(r^2 Y^2+e^2)+r^2eY(1-2\gamma)+2e(e^2+11r^2 Y^2).
\]
As long as $a_{1,2,3}$ remain positive, and $a_1a_2-a_3>0$, the steady state $E$ will remain stable in the odd subspace. However, provided $a_{1,2,3}$ remain positive, but $a_1a_2-a_3$ changes its sign, the steady state $E$ would become unstable through a Hopf bifurcation in the odd subspace. If any of the $a_1$ or $a_2$ become negative, this would mean one of the eigenvalues going through zero along the real axis implying a steady-state bifurcation and the loss of stability of the steady state $E$.\hfill$\blacksquare$\\

\vspace{0.2cm}

\noindent{\bf Proof of Theorem 2.} As it has already been explained, the steady states $E_{1,2,3,4}$ all lie on the same group orbit. In the light of equivariance of the system, this implies that all these states have the same stability type, and therefore it is sufficient to consider just one of them, for example, $E_1$. The Jacobian of linearisation near $E_1$ is given by
\[
\begin{array}{l}
J_1=\\\\
\small{\left(
\begin{array}{cccccccccccc}
0&0&0&0&-r(1-\gamma)Y_1&0&0&0&-r\gamma Y_1&0&0&0\\
0&r(1-\gamma W_1)-1&0&0&0&0&0&0&0&0&0&0\\
0&0&r-1&0&0&0&0&0&0&0&0&0\\
0&0&0&r(1-\gamma W_1)-1&0&0&0&0&0&0&0&0\\
r(1-Z_1)&0&0&0&-rY_1-e&0&0&0&0&0&0&0\\
0&r&0&0&0&-e&0&0&0&0&0&0\\
0&0&r&0&0&0&-e&0&0&0&0&0\\
0&0&0&r&0&0&0&-e&0&0&0&0\\
r(1-W_1)&r(1-W_1)&0&r(1-W_1)&0&0&0&0&-rY_1-e&0&0&0\\
r(1-W_1)&r(1-W_1)&r(1-W_1)&0&0&0&0&0&0&-rY_1-e&0&0\\
0&r&r&r&0&0&0&0&0&0&-e&0\\
r(1-W_1)&0&r(1-W_1)&r(1-W_1)&0&0&0&0&0&0&0&-rY_1-e
\end{array}
\right)},
\end{array}
\]
with the characteristic equation for eigenvalues that can be factorized as follows
\begin{eqnarray*}
&&(\lambda+e)^4[\lambda-(r-1)](rY_1+e+\lambda)^3[(e+rY_1)\lambda^2+(e+rY_1)^2\lambda+r^2eY_1]\times\\
&&\Big[\lambda+\frac{r^2Y_1(1-\gamma)+e(r-1)+rY_1}{e+rY_1}\Big]^2=0.
\end{eqnarray*}
It follows from this characteristic equation that one of the eigenvalues is $\lambda=r-1$, and since the steady state $E_1$ is only feasible for $r>1$, this implies that the steady state $E_1$ is unstable, and the same conclusion holds for $E_{2}$, $E_3$ and $E_4$.\hfill$\blacksquare$\\

\vspace{0.2cm}

\noindent{\bf Proof of Theorem 3.} Using the same approach as in the proof of {\bf Theorem 2}, due to equivariance of the system and the fact that within each cluster all the steady states lie on the same group orbit, it follows that for the analysis of stability of these steady states it is sufficient to consider one representative from each cluster, for instance, $E_{12}$ and $E_{13}$.

The Jacobian of linearisation near the steady state $E_{12}$ is given by
\[
\begin{array}{l}
J_{12}=\\\\
\small{\left(
\begin{array}{cccccccccccc}
0&0&0&0&-r(1-\gamma)Y_2&0&0&0&-r\gamma Y_2&0&0&0\\
0&0&0&0&0&-r(1-\gamma)Y_2&0&0&0&-r\gamma Y_2&0&0\\
0&0&r(1-\gamma W_{21})-1&0&0&0&0&0&0&0&0&0\\
0&0&0&r(1-\gamma W_{21})-1&0&0&0&0&0&0&0&0\\
r(1-Z_2)&0&0&0&-rY_2-e&0&0&0&0&0&0&0\\
0&r(1-Z_2)&0&0&0&-rY_2-e&0&0&0&0&0&0\\
0&0&r&0&0&0&-e&0&0&0&0&0\\
0&0&0&r&0&0&0&-e&0&0&0&0\\
r(1-W_{22})&r(1-W_{22})&0&r(1-W_{22})&0&0&0&0&-2rY_2-e&0&0&0\\
r(1-W_{22})&r(1-W_{22})&r(1-W_{22})&0&0&0&0&0&0&-2rY_2-e&0&0\\
0&r(1-W_{21})&r(1-W_{21})&r(1-W_{21})&0&0&0&0&0&0&-rY_2-e&0\\
r(1-W_{21})&0&r(1-W_{21})&r(1-W_{21})&0&0&0&0&0&0&0&-rY_2-e
\end{array}
\right)},
\end{array}
\]
The associated characteristic equation for eigenvalues has the form
\begin{equation}\label{ch_eq_12}
\begin{array}{l}
\displaystyle{(\lambda+e)^2(\lambda+e+rY_2)^2(\lambda+e+2rY_2)\left(\lambda-\frac{r^2Y_2(1-\gamma)+e(r-1)-rY_2}{rY_2+e}\right)^2\times}\\
\displaystyle{\left[(rY_2+e)x^2+(e^2+2Yer+Y^2r^2)x+r^2eY(1-\gamma)\right]P_3(\lambda)=0,}
\end{array}
\end{equation}
where $P_3(\lambda)$ is a third degree polynomial in $\lambda$
\[
P_3(\lambda)=\lambda^3+a_1\lambda^2+a_2\lambda+a_3,
\]
with
\[
\begin{array}{l}
a_1=3rY_2+2e>0\\\\
\displaystyle{a_2=\frac{2Y_2^2er^3(1+6Y_2)+Y_2e^2r[r(1+13Y_2)+r\gamma+6e]+4r^4Y_2^4}{(rY_2+e)(2rY_2+e)}>0,}\\\\
\displaystyle{a_3=\frac{r^2eY_2[2Y_2^2r^2(2-\gamma)+e^2(1+\gamma)+4\gamma e r]}{(rY_2+e)(2rY_2+e)}>0.}
\end{array}
\]
Computing $a_1a_2-a_3$ gives
\[
a_1a_2-a_3=reY_2[r(1+\gamma)+13rY_2+9e]+2(e^3+3r^3Y_2^3),
\]
which with the help of Routh-Hurwitz criterion (\ref{RHcon}) implies that all roots of $P_3(\lambda)$ lie in the left complex half-plane.

It follows that all the roots of the characteristic equation (\ref{ch_eq_12}) have negative real part except, possibly, an eigenvalue given by
\[
\lambda=\frac{r^2 Y_2(1-\gamma)+e(r-1)-rY_2}{e+rY_2}.
\]
Substituting the expression for $Y_2$ from (\ref{Y2def}), it can be shown that this eigenvalue crosses zero when $r=(1-2\gamma)/(1-\gamma)$ and $r=1$. Since $0\leq\gamma\leq 1$, and due to the fact that the steady state $E_{12}$ is only biologically plausible for $r>1$, it follows that stability of this steady state never changes as $r$ is varied irrespective of the value of $\gamma$, and, in fact, this steady state is always unstable.\\

\noindent In a similar way, the Jacobian of linearisation near the steady state $E_{13}$ has the form
\[
\begin{array}{l}
J_{13}=\\\\
\small{\left(
\begin{array}{cccccccccccc}
0&0&0&0&-r(1-\gamma)Y_3&0&0&0&-r\gamma Y_3&0&0&0\\
0&r(1-\gamma W_{32})-1&0&0&0&0&0&0&0&0&0&0\\
0&0&0&0&0&0&-r(1-\gamma)Y_3&0&0&0&-r\gamma Y_3&0\\
0&0&0&r(1-\gamma W_{32})-1&0&0&0&0&0&0&0&0\\
r(1-Z_3)&0&0&0&-rY_3-e&0&0&0&0&0&0&0\\
0&r&0&0&0&-e&0&0&0&0&0&0\\
0&0&r(1-Z_3)&0&0&0&-rY_3-e&0&0&0&0&0\\
0&0&0&r&0&0&0&-e&0&0&0&0\\
r(1-W_{31})&r(1-W_{31})&0&r(1-W_{31})&0&0&0&0&-rY_3-e&0&0&0\\
r(1-W_{33})&r(1-W_{32})&r(1-W_{32})&0&0&0&0&0&0&-2rY_3-e&0&0\\
0&(1-W_{31})r&r(1-W_{31})&r(1-W_{31})&0&0&0&0&0&0&-rY_3-e&0\\
r(1-W_{32})&0&r(1-W_{32})&r(1-W_{32})&0&0&0&0&0&0&0&-2rY_3-e
\end{array}
\right)},
\end{array}
\]
with the associated characteristic equation
\[
\begin{array}{l}
\displaystyle{(\lambda+e)^2(\lambda+e+rY_3)^2(\lambda+e+2rY_3)^2\left(\lambda-\frac{2r^2Y_3(1-\gamma)+e(r-1)-2rY_3}{2rY_3+e}\right)^2\times}\\
\displaystyle{\left[(e+rY_3)\lambda^2+(e+rY_3)^2\lambda+r^2eY_3\right]^2=0.}
\end{array}
\]
All of the eigenvalues given by the roots of this characteristic equation have negative real part, except for
\[
\lambda=\frac{2r^2Y_3(1-\gamma)+e(r-1)-2rY_3}{2rY_3+e}=\frac{2r^2(1-\gamma)+r(2\gamma-3)+1}{2r-1}.
\]
Solving the equation $\lambda=0$ shows that the steady state $E_{13}$ is stable when
\[
r<\frac{1}{2(1-\gamma)},
\]
and unstable otherwise. In the light of the restriction $r>1$, the steady state $E_{13}$ can only be stable for $\gamma>1/2$.\hfill$\blacksquare$

\end{document}